\documentclass[pre,showpacs,amsmath,amssymb,superscriptaddress,floatfix]{revtex4}
\usepackage{graphicx,amsmath}
\usepackage{color}
\usepackage{colordvi}
\begin{document}
\title{ASYMPTOTIC ENERGY PROFILE OF A WAVEPACKET IN DISORDERED CHAINS}

\author{S. Lepri}
\affiliation{Istituto dei Sistemi Complessi, Consiglio Nazionale delle
Ricerche, via Madonna del piano 10, I-50019 Sesto Fiorentino, Italy}

\author{R. Schilling}
\affiliation{Institut f\"ur Physik, Johannes Gutenberg-Universit\"at
D-55099 Mainz, Germany}

\author{S. Aubry}
\affiliation{Laboratoire Le\'on Brillouin
CEA Saclay, 91191 Gif-sur-Yvette, France}
\affiliation{Max Planck Institute f\"ur Physik komplexer Systeme,
N\"{o}thnitzer Str. 38, D-01187 Dresden, Germany}

\date{\today}


\begin{abstract}
We investigate the long time behavior of a  wavepacket initially
localized at a single site $n_0$ in translationally invariant harmonic
and anharmonic chains with  random interactions. In the harmonic case,
the energy profile
$ \overline{\langle e_n(t)\rangle}$ averaged on time and disorder decays
for large $|n-n_0|$ as a power law 
$\overline{\langle e_n(t)\rangle}\approx  C|n-n_0|^{-\eta}$
where $\eta=5/2$ and $3/2$ for initial displacement and momentum
excitations, respectively. The prefactor $C$ depends on the probability
distribution of the harmonic coupling constants and diverges in the
limit of weak disorder. As a consequence, the 
moments $\langle m_{\nu}(t)\rangle$
of the energy distribution averaged with respect to disorder diverge
in time as $t^{\beta(\nu)}$ for $\nu \geq 2$,  where
$\beta=\nu+1-\eta$ for $\nu>\eta-1$.
Molecular dynamics simulations yield good agreement with these
theoretical predictions. Therefore, in this system, the second moment of the
wavepacket  diverges as a function of time despite the wavepacket is
not spreading. Thus, this only criteria often
considered earlier as proving the spreading of a wave packet, cannot
be considered as sufficient in any model.
The anharmonic case is investigated numerically. It is found  for
intermediate disorder, that the tail of the energy profile becomes
very close to those of the
harmonic case. For weak and strong disorder, our results suggest that
the crossover to the harmonic behavior occurs at much larger $|n-n_0|$
and larger time.
\end{abstract}

\maketitle

\section{Introduction}
There has been large activity for many years in the study of the
temporal evolution of an initially localized energy excitation in
various nonlinear systems, e.g.~the discrete, nonlinear
Schr\"odinger equation (DNLS)
\cite{Shepelyansky93,Molina98,Pikovsky08,Kopidakis08},
Fermi-Pasta-Ulam (FPU) \cite{Bourbonnais90,Zavt93,Leitner01,Snyder06}
and Klein-Gordon (KG) model \cite{Kopidakis08, FLO9} with both uniform
and random couplings.
In the latter case, the main interest is in the interplay of
anharmonicity (nonlinearity) and disorder which is not yet fully understood. For harmonic
one-dimensional disordered systems, all eigenmodes (called Anderson modes) of the
{\it infinite} system are known to be localized and form a complete basis.
Then a wave packet at time $t=0$  will remain localized at any time as a linear superposition of
Anderson modes of  the infinite chain. Whether or not this behavior changes qualitatively by
introduction of anharmonicity is highly debated and controversial
(see Refs. \cite{Pikovsky08,Kopidakis08,FLO9} and references
therein).

Since an analytical treatment of the time evolution of anharmonic systems with
disorder is extremely difficult, most investigations have been done by molecular
dynamics simulations. In the numerical studies, one typically follows the
wavepacket dynamics by monitoring quantities like the participation ratio 
$P(t)$ (a measure the localization at time $t$), and the time-dependent moments
$m_\nu(t)$ of the local energy $e_n(t)$ (see the definitions below). All this
measurements are hampered by statistical errors as well as finite size
and finite time effects. Even very long calculation times of, say, 10$^8$
microscopic time units (of order picoseconds) may not be entirely conclusive.
Indeed, one can never be sure whether the spreading of a wavepacket is complete
or only partial in the infinite-time limit. This issues are intimately related
to the  spontaneous self-trapping of energy (for example in the form of discrete
breathers) which is generic in most nonlinear systems. 

Independently on complete or incomplete
spreading, one might expect that the evolution of the wavepacket tails
should yield relevant information on the spreading process itself.
In such regions, the typical displacement becomes small enough such that
linear approximation of the forces becomes valid. This motivates the
investigation of the harmonic chain, as a first necessary step for an insight
of the nonlinear case. 
Despite the apparent simplicity of such a case there are still 
issues that have not been fully discussed in the literature. 
Let us
briefly review some of the main results known for this case.
Without disorder all eigenstates are extended and it is
well-known (see e.g. Ref.~\cite{dk}, and references therein) that
\begin{equation} \label{mom}
m_\nu(t)\sim  t^{\beta(\nu)} \quad
\end{equation}
with $\beta(2)=2$, i.e. the energy spreading is ballistic (note
that $\nu$ is not necessarily an integer). Introducing disorder
and/or anharmonicities, this energy transport is changed and may
be superdiffusive ($\beta(2)>1$), diffusive ($\beta(2)=1$) or
subdiffusive ($\beta(2)<1$), or it could become logarithmic or
disappear ($\beta(2)=0$). If
the initial excitation is at site zero with amplitude $u_0(0)$,
then the disorder averaged propagator $\langle u_n(t)
\rangle$ is one of the basic quantities. Although $\langle u_0(t)
\rangle$ for $t \rightarrow \infty$ is known analytically for
different classes of disorder \cite{ABSO81}, 
much less is known for $n\neq 0$. Approximating the Anderson modes by
plane waves with exponentially decaying amplitudes, it has been shown
in Ref.~\cite{Zavt93} that

\begin{equation} \label{approx}
\langle u _n (t) \rangle \equiv \frac{1}{2(\pi \xi_0 |n|)^{1/2}} \exp [- \frac{(|n|-ct)^2}{4 \xi_0 |n|}]
\end{equation}
for $|n| \rightarrow \infty$ and $t\rightarrow \infty$. Here, $c$ is the sound velocity and $\xi_0$
a measure of the localization length.
Eq.(\ref{approx}) shows  for $t\rightarrow \infty$ there are two humps which propagate
ballistically at  the sound velocity $c$, but with an amplitude which decays as $1/\sqrt{t}$.
Within its co-moving frame , these humps spread as for normal diffusion. Another approach
for calculating $\langle u _n (t) \rangle$ is  to use a scaling hypothesis \cite{ABSO81}

\begin{equation} \label{scal}
\langle \tilde{u} _n (\omega) \rangle = \langle \tilde{u}_0
(\omega) \rangle F(n/ \xi(\omega)) \quad , \quad \omega
\rightarrow 0
\end{equation}
for the Laplace transform of $\langle u_n (t) \rangle)$ for
$\omega \rightarrow 0$. A similar Ansatz can be made for $\langle
u_n(t) \rangle$ \cite{AH78,AB79,RR80}. Here, $\xi(\omega)$ denotes
a localization length.

In this paper we investigate the energy profile
$\overline{\langle e_n(t)\rangle}$ averaged on time and disorder
of a wavepacket originating an initially localized excitation.
We demonstrate that it asymptotically decays as a power law in 
space. Thus, the wavepacket remains localized only weakly
while its moments appear to diverge in time. This result, 
which, to the best of our knowledge, has not been reported
previously, must be taken into account expecially when 
attacking more difficult nonlinear case. Indead, some 
numerical results for the anharmonic
chain (a FPU model) will be critically analyzed on 
the basis of the results on the harmonic one.

The outline of our paper is as follows. In Section II we will
introduce the harmonic model, rephrase some of its well-known
properties, define the local energy $e_n(t)$ and give some
information on our numerical approach. A virial theorem for the
time averaged local kinetic and potential energy will be proven in
Section III. It will be applied in this section for the analytical
calculation of the time and disorder average of $e_n(t)$. The corresponding
analytical result will be compared with the numerical one.
Furthermore we will investigate the moments $m_\nu(t)$ of the
local energy $\langle e_n(t) \rangle$. The influence of
anharmonicity on $e _n(t)$ will be numerically studied in Section
IV, and the final Section V contains a summary and some
conclusions.

\section{THE DISORDERED HARMONIC CHAIN}

\subsection{Property of the Anderson modes}

As motivated above we  investigate the classical dynamics of a
disordered harmonic chain with lattice constant $a$ which is
invariant under translations. Its classical Hamiltonian  reads:
\begin{equation} \label{ham}
H=\sum\limits_{n} \,
\left[\frac{p_n^2}{2m} + \frac{1}{2}
K_n(u_{n+1} - u_n)^2 \right]\quad .
\end{equation}
Here, $u_n$ is the displacement of the particle at site $n$, $p_n$ the
corresponding conjugate momentum, $m$ the particle's mass, and
$K_n$ the random coupling constants between nearest neighbors.
The $K_n$ are independent random variables,
identically distributed with some probability distribution $p(K)$.
Stability requires all $K_n$ to be positive. In our numerical
approach, the system is finite with $N$ particles and with free
ends, i.e. $K_{\pm N/2}=0$. Otherwise, we shall perform analytical
calculations in the thermodynamic limit $N \rightarrow \infty$
where the choice of the boundary conditions does not matter.  The
equations of motion are
\begin{equation} m \ddot u_n =K_n(u_{n+1}-u_n) -
K_{n-1}(u_n-u_{n-1}). \label{hchain}
\end{equation}

The general solution of Eqs.~(\ref{hchain}) with initial
conditions $u_n(0) = u_n$, $\dot{u}_n(0) = \dot{u}_n$ is given by
\begin{equation} \label{com}
u_n(t) =U_0+\dot{U}_0 t + \upsilon_n(t)
\end{equation}
where
\begin{equation}
\label{rel} \upsilon_n(t)=\sum\limits_{n'} \Big[u_{n'}
\Big(\sum\limits_{\nu\neq 0} Q^{(\nu)}_n Q^{(\nu)}_{n'} \cos
\omega_\nu t \Big) + \dot{u}_{n'} \Big(\sum\limits_{\nu\neq 0}
\frac{1}{\omega_\nu} Q_n^{(\nu)} Q_{n'}^{(\nu)} \sin \omega_\nu
t\Big) \Big] \quad .
\end{equation}
and $U_0=\sum\limits_n u_n/N$, $\dot{U}_0=\sum\limits_n
\dot{u}_n/N$ are the position and velocity, respectively, of the
center of mass of the  whole chain.

The eigenmodes $Q_n^{(\nu)}$ with eigenfrequency $\omega_\nu$ can
be chosen as real  with indices $\nu$ in a countable set. They
satisfy
\begin{equation}
\label{eq13} K_n(Q_n^{(\nu)}-Q_{n+1}^{(\nu)}) + K_{n-1}
(Q_n^{(\nu)}-Q_{n-1}^{(\nu)})= m \omega_\nu^2 Q_n^{(\nu)}.
\end{equation}
 and they can be normalized, except the uniform  eigenmode
 $Q_n^{(0)}\equiv1$ with  $\omega_0=0$   which is extended and
 cannot be normalized for the infinite system.
 For any size $N$ of a finite system, the translation invariance of the model implies that
$\quad Q_n^{(0)} = \frac {1}{\sqrt{N}}$ is an eigenmode with
eigenfrequency $\omega_0=0$.
In the limit of an infinite system, all eigenmodes are localized,
except this single zero-frequency mode which is extended. However,
nothing changes in the problem when choosing  the center of mass
of the whole system  immobile at $U_0=0$ with $\dot{U}_0=0$.
Though the eigenspectrum is discrete for the infinite system, it
is dense. The corresponding density of states
\begin{equation} \label{dos}
g(\omega) = \lim\limits_{N \rightarrow \infty} \, \frac{1}{N} \,
\sum\limits_{\nu=1}^{N-1} \delta(\omega-\omega_\nu)
\end{equation}
is a smooth function which is known \cite{selfav} to be 
self-averaging, i.e. it is independent on the disorder
realization with probability one.
Moreover, in the small frequency limit, $\omega \to 0$, we have
\cite{MI70,Ishii73}
\begin{equation}
  g(\omega) \simeq  \frac{\sqrt{m\langle K^{-1} \rangle}} \pi \quad. \qquad
\label{iomega}
\end{equation}

The localized eigenmodes,  decay exponentially with a localization length \cite{MI70,Ishii73}
\begin{equation} \label{cor}
\xi_\nu = \xi (\omega_\nu) \cong \frac{8 \langle K^{-1}
\rangle/m}{\langle K^{-2} \rangle - \langle K^{-1} \rangle^2} \,
\omega^{-2}_\nu \, a \quad , \quad \omega _\nu \rightarrow 0
\end{equation}
which diverges at the lower ``band'' edge at $\omega_0=0$.

Then, if the chain is finite with length $L$, there is a frequency $\omega_L$
such that the localization length equals the system size, i.e $\xi(\omega_L)=L=a N$.
Consequently, only the eigenmodes with frequency
$\omega_\nu > \omega_L$ can be considered as well  localized inside the finite system.
while the remaining modes where
$\omega_\nu < \omega_L$ extend over the whole finite system.
 Their number which is of order of $\sqrt{N}$ goes to infinity in the limit of an infinite system
 despite their relative weight for $N \rightarrow \infty$ goes to zero as $1 / \sqrt{N}$.
 As a result,  they still play a role for transport quantities, like the energy diffusion constant
\cite{Zavt93,DKP89,dk} or the thermal conductivity \cite{LLP03}.
Actually, those relatively extended modes behave like acoustic modes whose effective sound velocity is
\begin{equation}
c \; =\; \sqrt{\frac{\langle K^{-1} \rangle^{-1}}{m}} \, \, a
\label{vel}
\end{equation}

Although these results were originally proven for a chain with
mass disorder they also hold for our model. Indeed, letting
$y_n=({u_{n+1}-u_n})/K_n$, Eq.~(\ref{hchain}) is mapped onto the
eigenequation with mass disorder. This property has already been
used above since the mass average $\langle m \rangle$ has been
replaced by $\langle K^{-1} \rangle$.

\subsection{Local energy and local virial theorem}

We define the local energy:
\[
e_n(t)=e^{(\rm kin)}_n (t) + e^{(\rm pot)}_n(t)
\]
with kinetic and potential parts
\begin{equation} \label{kin}
e^{(\rm kin)} _n (t) =\frac{m}{2} (\dot{u}_n(t))^2
\end{equation}
and
\begin{eqnarray} \label{pot}
e_n^{(\rm pot)}(t)= &&\frac{1}{2} K_{n-1} [u_n(t)-u_{n-1} (t)]
u_n(t)
-\nonumber\\
&& \frac{1}{2} K_n [u_{n+1} (t) - u_n(t)] u_n (t)
\end{eqnarray}
respectively. Then,  $\sum\limits_{n} \, e^{(\rm pot)}_n$
equals the total potential energy in Eq.~(\ref{ham}). We will
investigate the energy profile for a displacement excitation:

\begin{equation} \label{disex}
u_n(0) = A \delta_{n,n_0} \quad , \quad p_n(0) \equiv0
\end{equation}

and a momentum excitation:

\begin{equation} \label{momex}
u_n(0) \equiv 0 \quad , \quad p_n(0) = B\, \delta_{n,n_0} \quad .
\end{equation}

For calculating numerically $\{u_n(t)\}$  we considered the example of a
uniform and uncorrelated distribution of random couplings $K_n$  with probability
distribution
\begin{equation}
p(K)=
\begin{cases} \frac{1}{k(R-1)} & \text{if $k \le K \le Rk$,}
\\
0 &\text{otherwise.}
\end{cases}
\label{uniform}
\end{equation}
where  of course $R\ge 1$.

To explore the role of different disorder strengths we fixed $k=1$
and took different $R$. The choice of units is such that $m=1$ and
$a=1$. Note that with this particular choices the effective sound
velocity, Eq.~(\ref{vel}), is $c \; =\; \sqrt{\frac{R-1}{\ln R}}$.

Microcanonical simulations were performed for typically $N=8192$
particles with fourth order symplectic algorithm \cite{MA92}, with
typical time step $5 \times \, 10^{-3}$ or less. Although the
choice of the initial conditions, Eqs.~(\ref{disex}) and
~(\ref{momex}), implies $U_0=A/N$, $\dot{U}_0=0$ and $U_0=0$,
$\dot{U}_0=B/N$, respectively, these nonzero quantities are rather
small, since $A$ and $B$ are of order one and $N \gg 1$.

In our numerical experiments, we avoid that
the wavepacket reaches the chain boundaries
which may generate spurious finite-size effects
(reflexions etc.). Thus, one should restrict the maximum
simulation time $t_{\rm sim}$ to be
smaller than $t_{\rm max}\sim N/c$ where $c$ is the sound  velocity.

\begin{figure}[ht]
\begin{center}\leavevmode
\includegraphics[width=0.45\textwidth,clip]{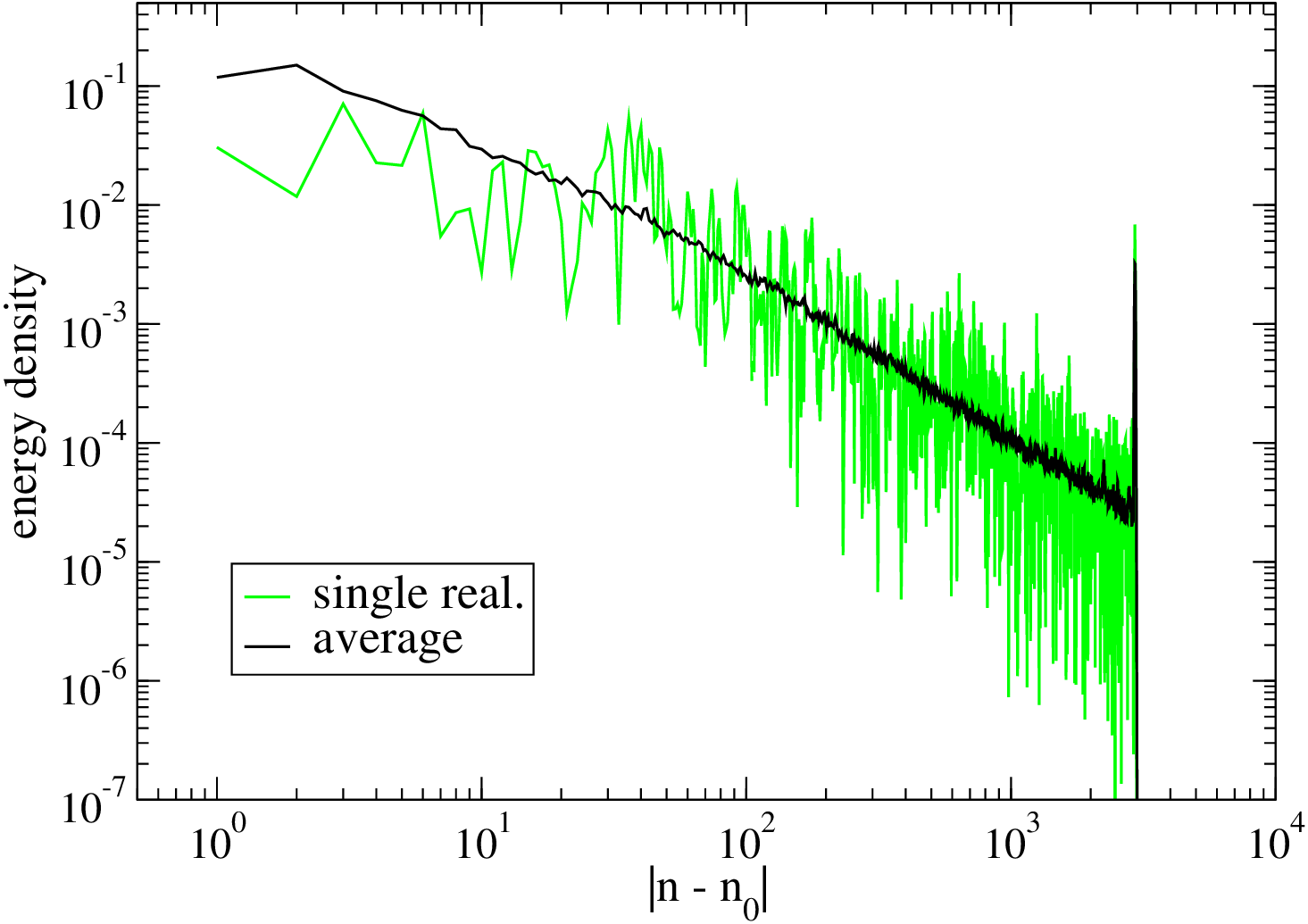}
\end{center}
\caption{(Color online) Energy profile $e_n(t)$ at $t=2000$ for a
momentum excitation with $B=2.0$, $N=8192$ particles, $R=4$ and
for a single realization of disorder (green line) and averaged
over $2\times 10^3$ realizations (black line).} \label{fig1}
\end{figure}

We also fixed $n_0=-N/2+1$ for extending the spatial range of our system,
so that one simulates the wavepacket
propagation in a semi-infinite medium \cite{Snyder06}. Some runs
with $n_0=0$ where also performed, yielding similar results.
Figure~\ref{fig1} shows the numerical profile $e_n(t=2000)$ for a
momentum excitation with $B=2.0$. The result for a {\it single}
realization of the disorder exhibits on the log-log-representation
strong fluctuations around an average, decaying linearly.
Averaging over a large enough number [$\mathcal O(10^3)$]
disorder realizations strongly reduces these
fluctuations and supports the linear dependence on $|n-n_0|$ on
the log-log scale. In the next Section we demonstrate that
this is indeed the case and compute analytically the exponents associated
with such power-lay decay.

The calculation of the time averaged energy profile will be
simplified by means of a local virial theorem, that will be proved
below. The well-known virial theorem \cite{LL82} relates the time
average of the {\it total} kinetic energy to the time average of
the virial. The virial \cite{LL82} involves the gradient of the
{\it total} potential energy. If the potential is harmonic this
theorem implies equality between the time averaged {\it total}
kinetic and {\it total} potential energy. In this subsection we
will prove that this relationship also holds for the time averaged
{\it local} kinetic and {\it local} potential energy, defined by
Eqs.~(\ref{kin}) and~(\ref{pot}), respectively.

The time average of a function $f(t)$ is defined by
\begin{equation} \label{timav}
\overline{f(t)}= \lim\limits_{T \rightarrow \infty} \frac{1}{T}
\int\limits_0^T \, dt \, f(t) \quad.
\end{equation}
Substitution of the general solution $u_n(t)$ of
Eqs.~(\ref{com}),~(\ref{rel}) into Eq.~(\ref{kin}) and taking into
account\\
\begin{eqnarray} \label{trigav}
&&\overline{\cos \omega_\nu t \cos \omega_{\nu'} t}=\frac{1}{2} \,
\delta_{ \nu \nu'}\nonumber\\
&& \overline{\sin \omega_\nu t \sin \omega_{\nu'} t} = \frac{1}{2}
\, \delta_{\nu \nu'}\nonumber\\
&& \overline{\sin \omega_\nu t \cos \omega_{\nu'} t} =0 \quad .
\end{eqnarray}
yields
\begin{equation} \label{potav}
\overline{e_n^{(\rm kin)} (t)} =\frac{m}{2} \dot{U}_0^2+
\frac{m}{4} \sum\limits_{\nu\neq 0} (Q^{(\nu)}_n)^2
\Big[\omega^2_\nu \Big(\sum\limits_{n'} u_{n'}
Q_{n'}^{(\nu)}\Big)^2 + \Big(\sum\limits_{n'} \dot{u}_{n'}
Q^{(\nu)}_{n'} \Big)^2 \Big] \quad .
\end{equation}
Note that the sum over $\nu$ remains discrete and cannot be
replaced by an integral in the limit of an infinite system.

With our definition Eq.~(\ref{pot})  of $e_n^{(\rm pot)}$,  and
using Eq.~(\ref{hchain}), we obtain:
\begin{equation} \label{epot2}
e_n^{(\rm pot)} (t)=-\frac{m}{2} \ddot{u}_n(t) u_n(t) \quad .
\end{equation}

Substituting $u_n(t)$ from Eqs.~(\ref{com}),~(\ref{rel}), and
since  $\upsilon_n(t)$ and $\dot{\upsilon}_n(t)$ has to remain
bounded at all times for any initially localized wavepacket (with
finite energy), yields
\begin{equation} \label{vir}
\overline{e_n^{(\rm pot)} (t)} = \overline{e_n^{(\rm kin)} (t)}
\quad ,
\end{equation}
for all $n$ and arbitrary initial conditions with finite energy in case that the
center of mass has been chosen immobile ($ \dot{U}_0 =0$).\\

\section{ENERGY PROFILE: HARMONIC CASE}

\subsection{Energy profile}
Without restricting generality we choose $m=1$ and $a=1$. Let us
discuss first the case of a displacement excitation for a given
disorder realization. In this case, we obtain from Eqs~(\ref{com})
and ~(\ref{rel}) for $A=1$ and $U_0=0$, $\dot{U}_0=0$

\begin{eqnarray}\label{eq20}
u_n(t) &=& \sum \limits _{\nu} Q_n^{(\nu)}
Q_{n_0}^{(\nu)} \cos \omega_\nu t\nonumber \\
\dot{u}_n(t) &=& - \sum \limits _{\nu \neq 0} \omega_\nu
Q_n^{(\nu)} Q_{n_0}^{(\nu)} \sin \omega_\nu t
\end{eqnarray}
and therefore
\begin{eqnarray}
\label{ekint} && e^{({\rm kin})}_n(t) =\frac{1}{2}
\sum\limits_{\nu,\nu'\neq 0} \, \omega_\nu \omega_{\nu'} \,
Q^{(\nu)}_n Q^{(\nu)}_{n_0} Q^{(\nu')}_n Q^{(\nu')}_{n_0} \sin
\omega_\nu t\,\sin \omega_{\nu'}t \quad .
\end{eqnarray}

Let us discuss first the qualitative $t$-dependence
of $e_n^{(\rm kin)}(t)$. We will explain how the spectral
properties govern its time dependence. Particulary we show that
this quantity which is not averaged over time and/or disorder does
not decay for $n \rightarrow \infty$ and/or $t \rightarrow \infty$. 
Since the eigenspectrum of the infinite random system is
discrete with a countable basis of localized eigenstates
$\{Q^{(\nu)}_n\}$, $u_n(t)$ has been expanded in this basis (see
Eq.~(\ref{rel}) and Eq.~(\ref{eq20})). This expansion is
actually an  absolutely convergent series of cosine functions of
time because
$$|\sum_{\nu} Q_{n_0}^{(\nu)} Q_{n}^{(\nu)}| \leq (\sum_{\nu} Q_{n_0}^{(\nu)~2})^{1/2} (\sum_{\nu} Q_{n}^{(\nu)~2})^{1/2}=1$$
Consequently, $u_n(t)$ is an almost periodic function in the sense of H. Bohr 
\cite{Bohr}.
An equivalent definition for such functions is that
for any arbitrarily small $\epsilon>0$, there is a monotone sequence of
$\tau_p$ ($p\in \mathcal{Z}$) (called pseudoperiods)  which is relatively dense (that is there exists $L$ such that $\tau_{p+1}-\tau_p < L$ for any $p$)  and such that for all $p$, $f(t)$ is periodic with period $\tau_p$ at the accuracy $\epsilon$
that is $|f(t+\tau_p)-f(t)| < \epsilon $ for any $p$ and for all $t$. As a consequence of this recurrence property,
an almost periodic function cannot go to zero for $t\rightarrow \pm \infty$.
The set of almost periodic functions is an algebra, that is linear combinations and products of
almost periodic functions are almost periodic functions, as well.

In our case, the  set of eigenfrequencies ${\omega_{\nu}}$ is bounded (since the support of the distribution
function $p(K)$ is compact) and thus
it is straightforward to show that the time derivative $\dot{u}_n(t)$ is also  an almost periodic function of time, and  the local kinetic energy  $ e^{({\rm kin})}_n(t)$  defined by Eq.~(\ref{kin}), as well.
$ e^{({\rm kin})}_n(t)$ from Eq.~(\ref{ekint}) can be decomposed into a time independent term and remaining
time dependent terms :

\begin{widetext}
\begin{eqnarray} \label{ekin}
&&e_n^{(kin)}(t) =\frac{1}{4} \sum \limits_{\nu \neq 0}
\omega_\nu^2 (Q^{(\nu)}_n Q^{(\nu)}_{n_0})^2 -\frac{1}{4}
\sum\limits_{\nu \neq 0} \omega^2_\nu (Q^{(\nu)}_n
Q_{n_0}^{(\nu)})^2 \cos 2 \omega_\nu t +\nonumber\\
&& +\frac{1}{4} \sum\limits_{\nu \neq \nu'}
\omega_\nu \omega_{\nu'} (Q_n^{(\nu)}Q^{(\nu)}_{n_0})
(Q^{(\nu')}_n Q^{(\nu')}_{n_0}) [\cos (\omega_\nu-\omega_{\nu'}) t
+ \cos (\omega_\nu + \omega_{\nu'})t] \quad.
\end{eqnarray}
\end{widetext}

Note that  for a finite chain without disorder,i.e. $K_n \equiv
K$, the first and second term on the r.h.s of Eq.(\ref{ekin}) are
of order $1/N$ since the eigenmodes are plane waves where
$Q_n^{(\nu)} \propto 1/\sqrt{N}$. Then $(Q_n^{(\nu)}
Q_{n_0}^{(\nu)})^2 \propto 1/N^2$, and there are only $N$ such
terms. Consequently, they will not contribute to $e_n^{(\rm kin)}
(t)$, in the limit $N \rightarrow \infty$ when the eigenspectrum
of the chain becomes absolutely continuous. In that case
$e_n^{(kin)}(t)$ can be represented by an integral which is a
Fourier transform of a smooth function and is obviously not an
almost periodic function. It decays to zero at infinite time as
expected from ballistic diffusion. This is not true in case of
disorder,  because each term in the series keeps a non vanishing
contribution for the infinite system and  $e^{({\rm kin})}_n(t)$
does not decay to zero at infinite time because it is almost
periodic.

%

However, in contrast to the ordered chain, $Q_n^{(\nu)} Q^{(\nu)}_{n_0}$ is not
 a smooth function of $\omega_\nu$, in case of disorder.
 The reason is that when the eigenspectrum is discrete,
arbitrarily small variations of $\omega_{\nu}$ may change the location of the corresponding
localized eigenstate by arbitrarily large distances.
Thus,  these  eigenstates $\{Q_n^{(\nu)}\}$ are not continuous functions of $\omega_{\nu}$
but depend on the disorder realization as well as  $e_n^{(\rm kin)} (t)$ and $e_n(t)$
(since they are obtained as discrete series  explicitly involving these eigenstates).
The consequence is that  those quantities  are not
self-averaging, as clearly demonstrated by Fig.~\ref{fig1} for $e_n^{(\rm kin)} (t)$.

\subsection{Disorder averaged profile}

Since $e_{n}^{(kin)}(t)$ is an almost periodic function of time, it is a stationary solution.
Its  time average drops all cosine terms in Eq. (\ref{ekin})  and keeps only the constant term,
i.e. we get $$\overline{e_n^{(kin)}(t)} =\frac{1}{4} \sum\limits_{\nu \neq 0}
\omega_\nu^2 (Q^{(\nu)}_n Q^{(\nu)}_{n_0})^2 .$$ An attempt to justify the use of the time averaged
quantity will be given below.
$\overline{e_n^{(kin)}(t)}$ and $\overline{e_n(t)}$ still depend on the disorder realization. Therefore it is reasonable to calculate the corresponding disorder averaged quantities, as well.  Despite they cannot be observed
for any single disorder realization, they give information on the general behavior of the profiles. Then we arrive at

\begin{equation}
\label{limprof} \langle\overline{e_n^{(\rm kin)} (t)}
 \rangle =\frac{1}{4} \langle \sum\limits_{\nu \neq 0} \omega^2_\nu
(Q^{(\nu)}_n Q^{(\nu)}_{n_0})^2 \rangle
\end{equation}
for the infinite system.

Note that, in the infinite system, the set of eigenvalues and
eigenvectors are discontinuous functions of the disorder
realization. Yet, according to Wegner \cite{Wegner75}, the
disorder average   $\langle F(\{Q^{(\omega)}_n\}) \rangle$ of an
arbitrary function $F(\{Q^{(\nu)}_n)\})$ of the eigenvectors can
be well defined as a smooth function of $\omega$  as a limit for
finite systems with size $N \rightarrow \infty$
$$\langle  F(\{Q^{(\omega)}_n)\}) \rangle g(\omega) \delta \omega
= \lim_{N\rightarrow + \infty}  \int
 \left( \frac{1}{N} \sum_{\omega < \omega_{\nu}<\omega +\delta \omega}  F(\{Q^{(\nu)}_n\})\right)  \prod_{n=1}^N p(K_n) d K_n$$

The sum in the integral is restricted  to eigenvalues
$\omega_{\nu}$ which belong to an interval  $[\omega,\omega
+\delta \omega]$ of small width $\delta \omega$ and $g(\omega)$ is
the density of states defined by Eq.~(\ref{dos}).


Then, we obtain from
Eq.~(\ref{limprof}) :

\begin{equation} \label{ekinav3}
\langle\overline{e_n^{(\rm kin)} (t)}\rangle =\frac{1}{4} \int\limits_0^{\infty}  d \omega \omega^2
g (\omega) \lim\limits_{N \rightarrow \infty}
 (N \langle (Q_n^{(\omega)} Q^{(\omega)}_{n_0})^2 \rangle) \quad .
 \end{equation}

Since $\overline{e_n^{(\rm pot)}(t)} = \overline{e_n^{(\rm
kin)}(t)}$, for $N=\infty$ and all realizations of $\{K_n \}$ the time and
disorder averaged energy profile is given by

\begin{equation} \label{avprof}
\langle \overline{e_n(t)} \rangle = 2 \langle \overline{e_n^{(\rm
kin)} (t)} \rangle \quad ,
\end{equation}

i.e.~the calculation of $\langle \overline{e_n (t)} \rangle$ is
reduced to that of $g(\omega)$ and the ``quadratic'' correlation
function $\langle (Q^{(\omega)}_n Q^{(\omega)}_{n_0})^2 \rangle$
for $N \rightarrow \infty$.  $\{Q^{(\omega)}_n\}$ is the solution of the eigenvalue
equation (\ref{eq13}) with $\omega_\nu$  replaced by $\omega$.

Before we come to the evaluation of the ``quadratic'' correlation function,
let us return to Eq.(\ref{ekin}). Making again use of the self-averaging of the
density of states we obtain for the second term on its r.h.s.:
$$-\frac{1}{4} \int\limits_0^{\infty} d \omega \,
\omega^2 g(\omega) \lim\limits_{N \rightarrow \infty} (N \langle
(Q_n^{(\omega)} Q^{(\omega)}_{n_0})^2 \rangle ) \cos 2 \omega t
\quad  .$$

Below, it will be shown that
$\lim\limits_{N \rightarrow \infty} (N \langle (Q^{(\omega)}_n
Q_{n_0}^{(\omega)})^2 \rangle )$ is a finite and smooth function
of $\omega$. Therefore, the disorder averaged second term will
converge to zero, for $t \rightarrow \infty$, due to $g(\omega)
\rightarrow g_0={\rm const.},$ for $\omega \rightarrow 0$. The
same property should hold for the disorder average of the square
bracket term in Eq.~(\ref{ekin}). With the  density of states
$g(\omega,\omega^{\prime})$ giving the joint distribution for
two eigenfrequencies  the disorder averaged  square bracket term
becomes a double integral  over $\omega$  and $\omega^{\prime}$.
 Although we do not have a rigorous proof, $\lim\limits_{N \rightarrow \infty} (N^2 \langle
(Q_n^{(\omega)} Q^{(\omega)}_{n_0}) (Q^{(\omega')}_n
Q_{n_0}^{(\omega')})\rangle )$ which is part of the integrand should be a finite and smooth
function of $\omega$  and $\omega^{\prime}$. Then, taking the limit $N \rightarrow \infty$
first, the square bracket term should converge to zero for $t\rightarrow \infty$.
If this is true the disorder averaged energy profile converges to an asymptotic profile for
$t \rightarrow \infty$ which is consistent with our numerical result. Indeed, the disorder averaged
profile in Fig.~\ref{fig1}  depends on $t$ only very weakly, for large $t$. In that case the asymptotic
profile equals the time averaged one.

Now we come back to the ``quadratic'' correlation function. Due to the disorder average it will
depend only on $|n-n_0|$. Since the Anderson modes are exponentially localized one expects that
this correlation function decays exponentially with $|n-n_0|$. To prove this we first present
a crude heuristic approach by assuming
\begin{equation}\label{heuristic-mode}
Q_n^{(\nu)} \approx \mathcal{N}_\nu \exp \left(- \frac{|n-n_\nu|}
{\xi_\nu}\right)
\end{equation}

where the ``center of mass'' of the Anderson mode $\nu$ is at
$n_\nu$, which is a random variable, depending on $\{K_n\}$.
$\mathcal{N}_{\nu}$ is a normalization constant. It should be 
remarked, that the envelope
of an Anderson mode $Q^{(\nu)}_n$ decays exponentially, but not
$Q^{(\nu)}_n$ itself. Therefore Eq.~(\ref{heuristic-mode}) is a
crude approximation neglecting sign changes of $Q^{(\nu)}_n$ with
$n$. Substituting $Q_n^{(\nu)}$ from Eq.(\ref{heuristic-mode})
into the ``quadratic'' correlation function and using :

\begin{eqnarray}\label{ergoprop}
\langle f(n_\nu)\rangle \approx \frac 1 N \sum \limits _{n_\nu=1}^N f(n_\nu)\quad ,
\end{eqnarray}

we get :

\begin{eqnarray}\label{approxcorrel}
\langle (Q_n^{(\nu)} Q^{(\nu)}_{n_0})^2 \rangle
\approx \frac 1 N \mathcal{N}_\nu ^4[\coth \frac {2}{\xi_\nu} + |n-n_0|] \exp [- \frac
{2}{\xi_\nu} |n-n_0|]
\end{eqnarray}

i.e. the ``quadratic'' correlation function decays exponentially.

For an analytical calculation of the ``quadratic'' correlation function in Eq.(\ref{ekinav3})
one can use the approach presented in Refs. \cite{Wegner75,Wegner81,Wegner85}. These authors prove that
the computation of the  correlation
functions  $\langle Q^{(\omega)}_n Q^{(\omega)}_{n_0} \rangle$ and $\langle
|Q^{(\omega)}_n||Q^{(\omega)}_{n_0} |\rangle$ for $|n-n_0| \rightarrow \infty$
is reduced to the solution of an eigenvalue problem
for an integral kernel. As a result, these correlation functions decay exponentially for large $|n-n_0|$
with an inverse localization length given by $-\ln|\lambda_{max}(\omega)|$. $|\lambda_{max}(\omega)| $ is the
largest absolute value of the eigenvalues of the kernel. It is smaller than one. Applying that approach
it follows for $|n-n_0| \rightarrow \infty$

\begin{equation} \label{quadcor}
\langle (Q^{(\omega)}_n Q^{(\omega)}_{n_0})^2 \rangle \cong \alpha
(\omega) \exp (-|n-n_0|/\xi_2 (\omega))
\end{equation}



with a correlation length $\xi_2 (\omega)$. We note that the
eigenvalue problem in form of an integral equation can only be
used to calculate correlation functions of the Anderson modes and
not directly to compute the energy profile itself. But the former
is needed (see Eq.~(\ref{ekinav3})) for the latter.

The correlation lengths (localization lengths) of the correlation
functions calculated in Refs. \cite{Wegner75,Wegner81,Wegner85}
and of the ``quadratic'' correlation function Eq.~(\ref{quadcor})
are different from each other and different from $\xi(\omega)$
(Eq.~(\ref{cor})), for finite $\omega$. But for $\omega
\rightarrow 0$ they exhibit the same divergence, i.e.~it is (see
Eq.~(\ref{cor})):
\begin{equation} \label{quadloc}
\xi_2 (\omega) \cong c_2 \omega^{-2} \quad , \quad \omega
\rightarrow 0
\end{equation}
with a positive constant $c_2$, depending on $p(K)$.

The pre-exponential factor $\alpha (\omega)$ can  be determined as
follows. Assuming that Eq.~(\ref{quadcor}) is valid for all
$|n-n_0|$, summation of the l.h.s. and r.h.s. of that equation and
accounting for the normalization $\sum\limits_n
(Q^{(\omega)}_n)^2=1$ for $\omega=\omega_\nu$ (remember that
$Q^{(\nu)}_n$ has been chosen as real) yields for $N \rightarrow
\infty$:

\begin{eqnarray} \label{pref}
\alpha (\omega) &\cong& \frac {1} {N \coth (1/\xi_2(\omega))} \nonumber \\
&\cong& \frac {\omega ^2}{Nc_2} \quad , \quad \omega \rightarrow 0
\quad .
\end{eqnarray}

In the last line, Eq.~(\ref{quadloc}) has been applied. With Eqs.~
(\ref{quadcor}), (\ref{pref}) and (\ref{ekinav3}), it follows from
Eq.~(\ref{avprof}):

\begin{equation}\label{avprof2}
\langle \overline{e_n(t)}\rangle \cong \frac{1}{2} \int \limits
_0^{\infty} d \omega \; g(\omega) \omega ^2 \frac
{\exp(-|n-n_0|/\xi_2(\omega))}{\coth (1/\xi_2(\omega))} \quad .
\end{equation}

The asymptotic $|n-n_0|$-dependence is governed by the
small-$\omega$ behavior of the integrand. From Eq.~(\ref{iomega})
we get

\begin{equation}\label{dos2}
g(\omega) = d I(\omega)/d \omega \cong \sqrt{\langle
K^{-1}\rangle}/\pi \quad , \quad \omega \rightarrow 0\;.
\end{equation}

Assuming that Eqs.~(\ref{quadloc}) and (\ref{dos2}) are valid for
all $\omega$ will not influence the asymptotic dependence of
$\langle \overline{e_n(t)}\rangle $ on $|n-n_0|$. Then we get from
Eq.~(\ref{avprof2}) for the infinite chain and a
\textit{displacement excitation}:

\begin{equation}\label{asym1}
\langle \overline{e_n(t)}\rangle \cong \frac {3} {16}
\sqrt{c_2^3\langle K^{-1}\rangle /\pi} \,\,\, |n-n_0|^{-5/2},
\,\,\, |n-n_0| \rightarrow \infty,
\end{equation}

i.e. the time and disorder averaged energy profile decays as a
power law in $|n-n_0|$ with an exponent $\eta = 5/2$.

So far we have discussed the energy profile for a displacement
excitation. The corresponding calculation for a momentum
excitation is similar. With the initial condition
Eq.~(\ref{momex}) and $B=1$, Eq.~(\ref{rel})leads to

\begin{equation}\label{udotmom}
\dot{u}_n(t) =  \sum \limits _{\nu =1}^{N-1}
Q_n^{(\nu)}Q_{n_0}^{(\nu)} \cos \omega _\nu t \quad \quad .
\end{equation}

Besides $\cos \omega _\nu t$ the main difference to $\dot{u}_n(t)$
for a displacement excitation (see Eq.~(\ref{eq20})) is the
absence of the prefactor $\omega _\nu$ of
$Q_n^{(\nu)}Q_{n_0}^{(\nu)}$. As a consequence one obtains

\begin{equation}\label{avprofmom}
\langle \overline{e_n(t)}\rangle \cong \frac{1}{2} \int \limits
_0^\infty d\omega g(\omega) \frac
{\exp(-|n-n_0|/\xi_2(\omega))}{\coth (1/\xi_2(\omega))}
\end{equation}

where $\omega ^2$ in Eq.~(\ref{avprof2}) is replaced by one. With
the same assumptions as above we obtain for the infinite chain and
a \textit{momentum excitation}:

\begin{equation}\label{asym2}
\langle \overline{e_n(t)}\rangle \cong \frac {1} {8}
\sqrt{c_2\langle K^{-1}\rangle /\pi} \,\,\, |n-n_0|^{-3/2}, \quad
\quad |n-n_0| \rightarrow \infty \quad.
\end{equation}

It is not surprising that we find a power law decay again. The
corresponding exponent is $\eta = 3/2$.

\begin{figure}[ht]
\begin{center}\leavevmode
\includegraphics[width=0.45\textwidth,clip]{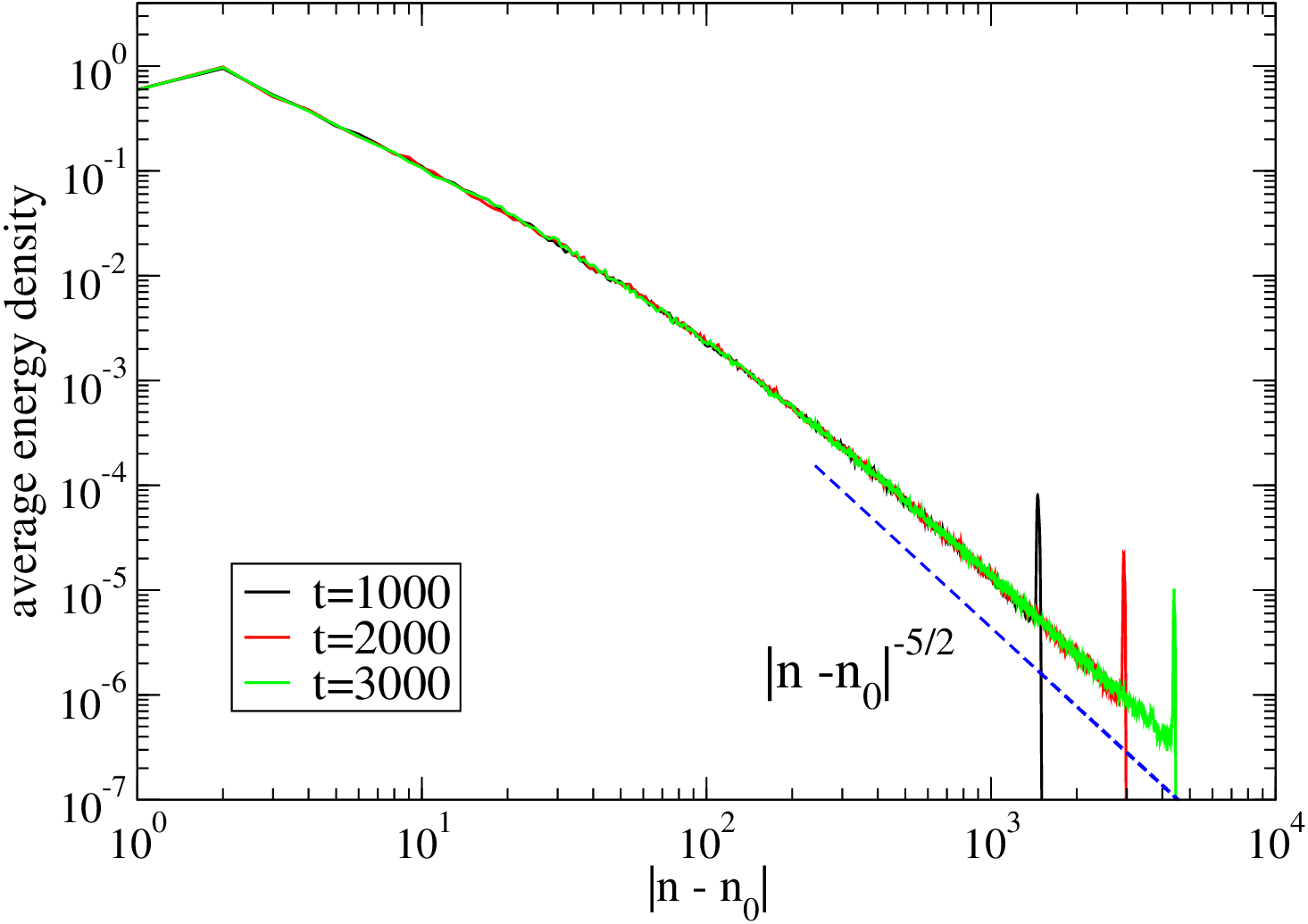}
\end{center}
\caption{(Color online) Energy profile at three different times
averaged over 10$^3$ realizations of the disorder with $R=4$ for
$N=8192$ particles and a displacent excitation with $A=2$.
The dashed line is the predicted
power law decay, Eq.~(\ref{asym1}).
The ballistic peaks propagate at a velocity $c=1.476$ in
agreement with the value $c=1.471...$ computed from Eq.~(\ref{vel}).
}\label{fig2}
\end{figure}

\begin{figure}[ht]
\begin{center}\leavevmode
\includegraphics[width=0.45\textwidth,clip]{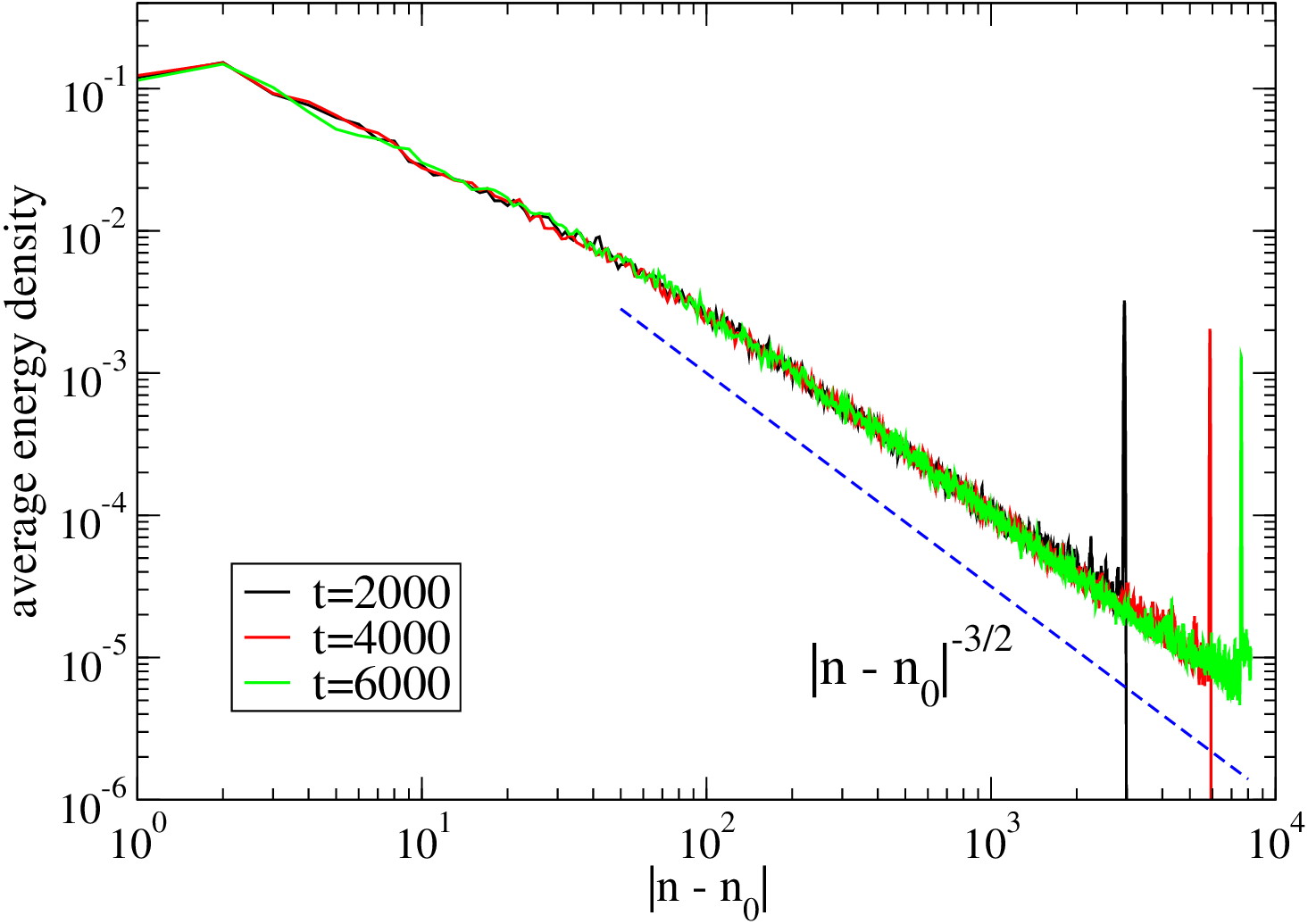}
\end{center}
\caption{(Color online) Same as Fig.~\ref{fig2} but for a momentum
excitation with $B=2$ and averaged over $2 \times 10 ^3$
realizations. The dashed line is the predicted power law,
Eq.~(\ref{asym2}).}\label{fig3}
\end{figure}

Figures~\ref{fig2} and \ref{fig3} report the numerical result for
the disorder averaged energy profile at different large times of a
displacement and momentum excitation, respectively. They clearly
demonstrate first that, the numerical result of the disorder
averaged energy profile becomes independent of $t$ for $t$ large
enough, and second, that it converges to a power law for large
$|n-n_0|$ with exponents predicted by the analytical calculation.
The three spikes at the $t$-dependent positions $n(t)$ are the
phonon fronts propagating with the effective sound velocity,
Eq.~(\ref{vel}).

There is a finite size effect for $N <\infty$, due to the
existence of extended states. For $\omega \alt \omega _L=
\theta/\sqrt{N}$, (with $\theta$ being a suitable constant
$\mathcal{O}(1)$), it is:
\begin{equation}\label{ext}
\langle (Q_n^{(\omega)} Q_{n_0}^{(\omega)} )^2 \rangle \approx
\frac {4} {N} \sin^2 (cqn) \sin ^2 (cqn_0)
\end{equation}
where $\omega \cong cq$. With $g(\omega) \approx g_0$ for the
density of extended states it is easy to estimate the contribution
of those to $\langle \overline{e_n(t)}\rangle$ in case of a
displacement excitation:

\begin{equation}\label{profext1}
\langle \overline{e_n(t)} \rangle ^{(\textrm{ext})} \approx \frac
{\theta^3g_0}{6}N^{-5/2}\quad \quad .
\end{equation}

If $1 \ll|n-n_0|\ll N$, then it is $\langle
\overline{e_n(t)}\rangle = \langle \overline{e_n(t)}\rangle
^{(\textrm{loc})} + \langle \overline{e_n(t)}\rangle
^{(\textrm{ext})} \cong \langle \overline{e_n(t)}\rangle
^{(\textrm{loc})} \sim |n-n_0|^{-5/2}$. For $|n-n_0| <N$ but
$|n-n_0|= \mathcal{O}(N)$ there is a crossover value
$|n-n_0|_{c.o.}$ depending on $\theta,g_0$ etc. such that

\begin{equation}\label{profext2}
\langle \overline{e_n(t)}\rangle ^{(\rm loc)} \cong \langle
\overline{e_n(t)}\rangle ^{{(\textrm{ext})}} \sim
N^{-5/2}\quad \quad .
\end{equation}

For $N=8192$ this contribution is of order 10$^{-10}$. The
corresponding contribution for a momentum excitation is
\begin{equation}\label{profext3}
\langle \overline{e_n(t)} \rangle ^{(\textrm{ext})} \sim N^{-3/2}
\quad ,
\end{equation}
being of order 10$^{-6}$ for $N=8192$.\\

\begin{figure}[ht]
\begin{center}\leavevmode
\includegraphics[width=0.45\textwidth,clip]{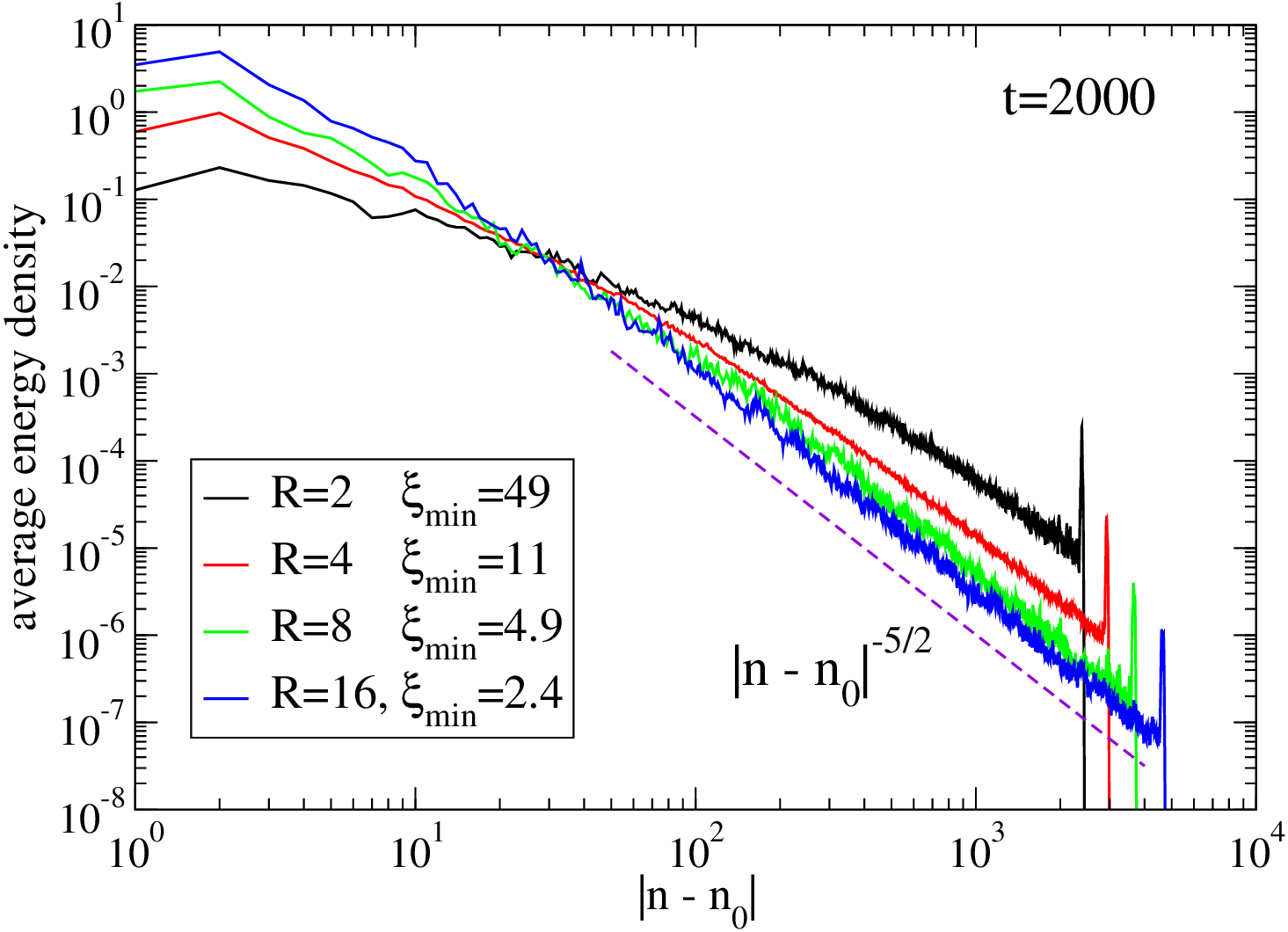}
\end{center}
\caption{(Color online) Disorder averaged energy profiles at
$t=2000$ for a displacement excitation with $A=2$ and increasing
disorder strengths (top to bottom). Other parameters as in
Fig.\ref{fig1}. The dashed line indicates the predicted power law
Eq.~(\ref{asym1})}\label{fig4}
\end{figure}

Figure~\ref{fig4} compares the energy profiles for different
strengths of the disorder, i.e. for various values of the
parameter $R$. We limited ourselves to the case of a displacement
excitation. The profiles display the same decay law. The cases
with stronger disorder attain the asymptotic profile at smaller
distances since in this case the localization lengths are shorter.
As seen from Figure~\ref{fig4}, the data are consistent with the
expectation that the asymptotic profile is reached for $|n-n_0|
\gg \xi_{\textrm{min}}$. The values of $\xi_{\textrm{min}}$ given
in that figure are a rough estimate of the shortest localization
length obtained by extrapolating the formula (\ref{cor}) at
$\omega = \omega_{max}=2c$, i.e. $\xi_{min} = \xi(\omega_{max})$.

The prefactor of both power laws, Eqs.~(\ref{asym1}) and
(\ref{asym2}) depends on the disorder as demonstrated by Figure 4.
It seems reasonable that $\xi_2 \cong \lambda \xi (\omega)$ for
$\omega \rightarrow 0$ with a positive parameter $\lambda$,
independent on $\omega$ and the disorder. Eqs.~(\ref{cor}) and
(\ref{quadloc}), together with this hypothesis, imply:

\begin{equation}\label{c2}
c_2= 8 \lambda \frac {\langle K^{-1}\rangle}{\langle K^{-2}
\rangle - \langle K^{-1}\rangle ^2} \quad .
\end{equation}

Again $m=1$ and $a=1$ has been used. For the uniform distribution
$p(K)$, Eq.~(\ref{uniform}), $\langle K^{-1}\rangle $ and $\langle
K^{-2}\rangle$ can easily be calculated. From this we obtain:

\begin{eqnarray} \label{c2R}
&& c_2 (R)= 8\lambda\frac{(R-1) R \ln R}{(R-1)^2-R(\ln R)^2} \nonumber\\
&& \cong 72 \lambda (R-1)^{-2} [1 + \mathcal O(R-1)]
\end{eqnarray}

Note, that $c_2(R)$ diverges in the no-disorder-limit
$R \rightarrow 1^+$, as it should be since only
extended states exist thereby. Accordingly, $\xi(\omega)$ should
become infinite for all $\omega$.

\begin{figure}[ht]
\begin{center}\leavevmode
\includegraphics[width=0.45\textwidth,clip]{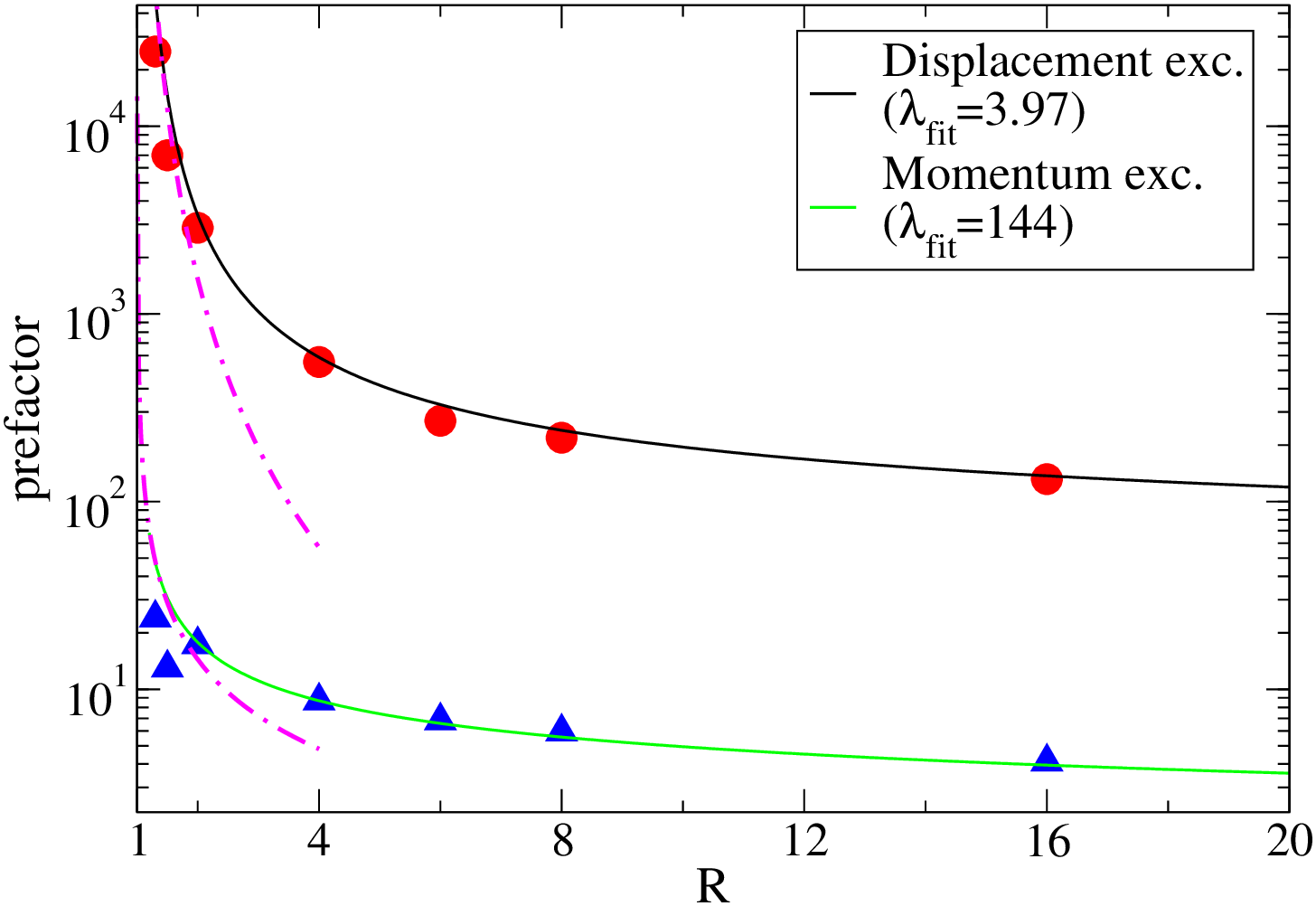}
\end{center}
\caption{(Color online) Dependence of the prefactors $C(R)$ on the
disorder strength $R$ for a displacement ($A=2$) and momentum ($B=2$) 
excitation: numerical (full circles and triangles
respectively) and
analytical result Eqs.~(\ref{asym1}), (\ref{asym2}) with formula
(\ref{c2R}) (solid lines). Dash-dotted lines are the expected
asymptotic behaviors for $R\to 1^+$, $(R-1)^{-3}$ and
$(R-1)^{-1}$, respectively. The fitting parameter $\lambda_{\rm
fit}$ is given in the legend. }\label{fig5}
\end{figure}

Introducing $c_2(R)$ from the first line of Eq.~(\ref{c2R}) and
$\langle K^{-1} \rangle (R) = (\ln R)/(R-1)$ into the prefactor
$C(R)$ of the power laws, Eqs.~(\ref{asym1}) and (\ref{asym2}),
leads to the $R$-dependence shown in Fig.~\ref{fig5} in case of a
displacement and a momentum excitation, respectively. The unknown
parameter $\lambda$ has been adjusted in order to fit the
numerical result for the prefactors. The latter are obtained from
the numerical data in Figures~\ref{fig2} and \ref{fig3}
extrapolating $\langle e_n (t) \rangle|n-n_0|^\eta$ at large
$|n-n_0|$.

The numerical and analytical result for the prefactor in case of a
displacement excitation agree satisfactorily, even for the
smallest value of $R=1.3$. Investigating the profile for even
smaller values is hampered for our finite chain by the increase of
the localization length with decreasing $(R-1)$. The same
agreement is also valid in case of the momentum excitation, except
for the two smallest $R$-values at 1.3 and 1.5. Eq.~(\ref{rel})
demonstrates that the weight of the low-lying Anderson modes for a
momentum excitation is by a factor $1/\omega_\nu$ higher than for
a displacement excitation. Since the localization length increases
with decreasing $\omega_\nu$, this could be the reason for the
``asymmetric'' behavior of $C(R)$ for both kind of excitations.
Indeed, we have observed that $\langle e_n(t) \rangle
|n-n_0|^{3/2}$ does not reach a stationary value, for e.g.
$R=1.3$. The strong deviation of the fit parameter $\lambda$ for
the displacement and momentum excitation may originate also from
this fact.

\subsection{Moments of the local energy}

\begin{figure}[ht]
\begin{center}\leavevmode
\includegraphics[width=0.45\textwidth,clip]{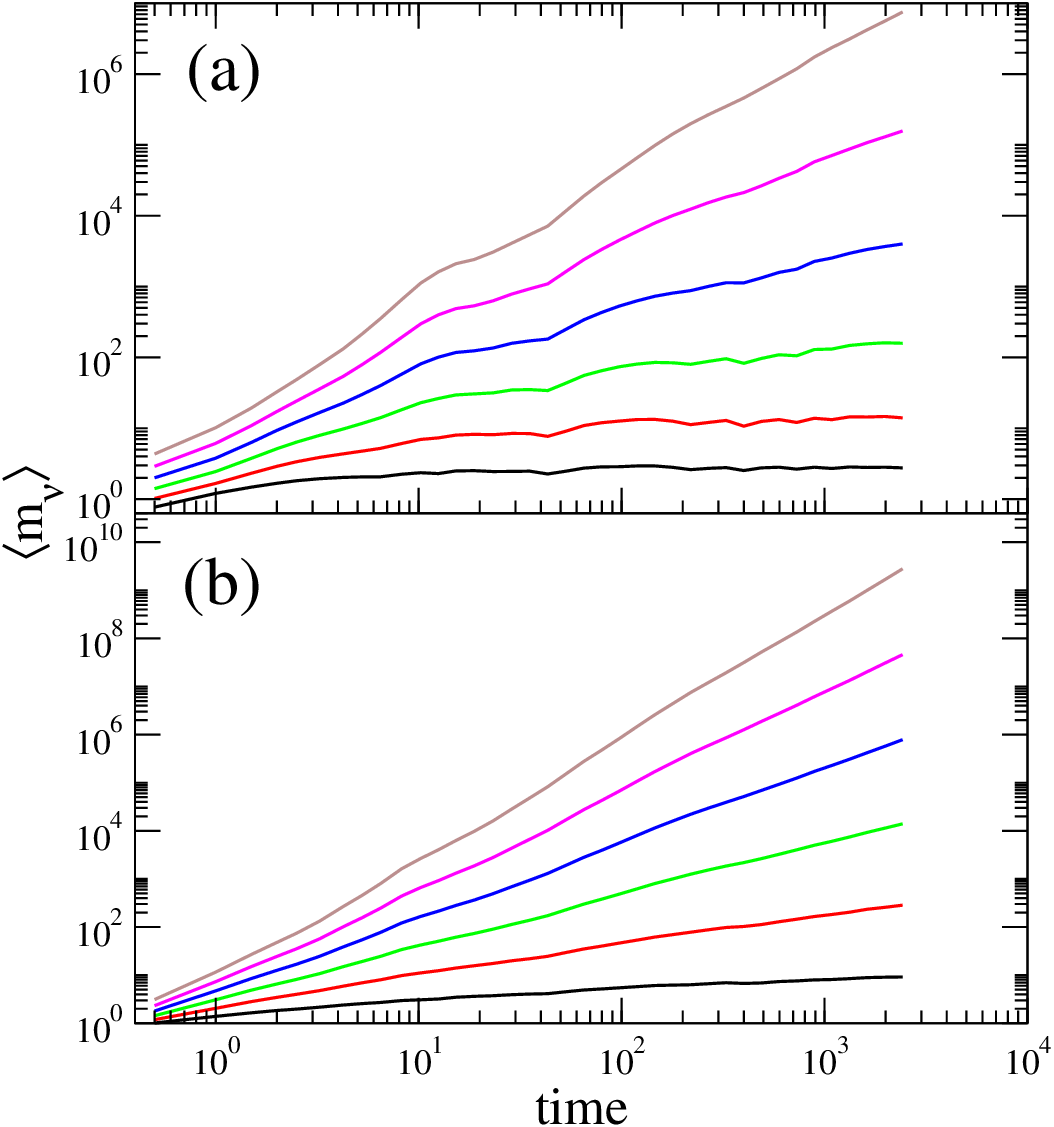}
\end{center}
\caption{(Color online) Evolution of
moments $\langle m_\nu(t) \rangle$  for
$\nu=0.5,1,1.5,2,2.5,3$ (bottom to top respectively)
for the harmonic chain $N=8192$, $R=6$.
Displacement (a) and momentum (b) excitation
($A=2$ and $B=2$ respectively). Averages are over $6 \times 10^3$
disorder realizations.
} \label{fig6}
\end{figure}

A customary way to describe wavepacket diffusion is to look at
time evolution of moments of the energy distribution that are
defined as
\begin{equation}
\label{momenta} m_\nu (t) \;=\; \frac{\sum_n \, |n-n_0|^\nu e_n
(t)}{\sum_n \,  e_n}
\end{equation}
(the denominator is clearly only a scale factor). Of particular
interest for a statistical characterization are the
disorder averaged moments $\langle m_\nu (t) \rangle$. Their
numerical result is shown in Figure \ref{fig6}. If one uses the
asymptotics Eqs.~(\ref{asym1}), (\ref{asym2}) and introduces a
cutoff of the sum in the numerator of Eq.~(\ref{momenta}) at the
ballistic distance $|n-n_0|=ct$ one obtains
\begin{equation}
\label{mnu} \langle m_\nu (t) \rangle \; \propto \; t^{\beta(\nu)}  ,
\quad \beta(\nu) \;=\; \left \{ \begin{array}{ll}
\nu+1-\eta & , \quad \nu > \eta-1 \\
0 & ,\quad \nu < \eta-1 \quad .
\end{array}\right.
\end{equation}

\begin{figure}[ht]
\begin{center}\leavevmode
\includegraphics[width=0.45\textwidth,clip]{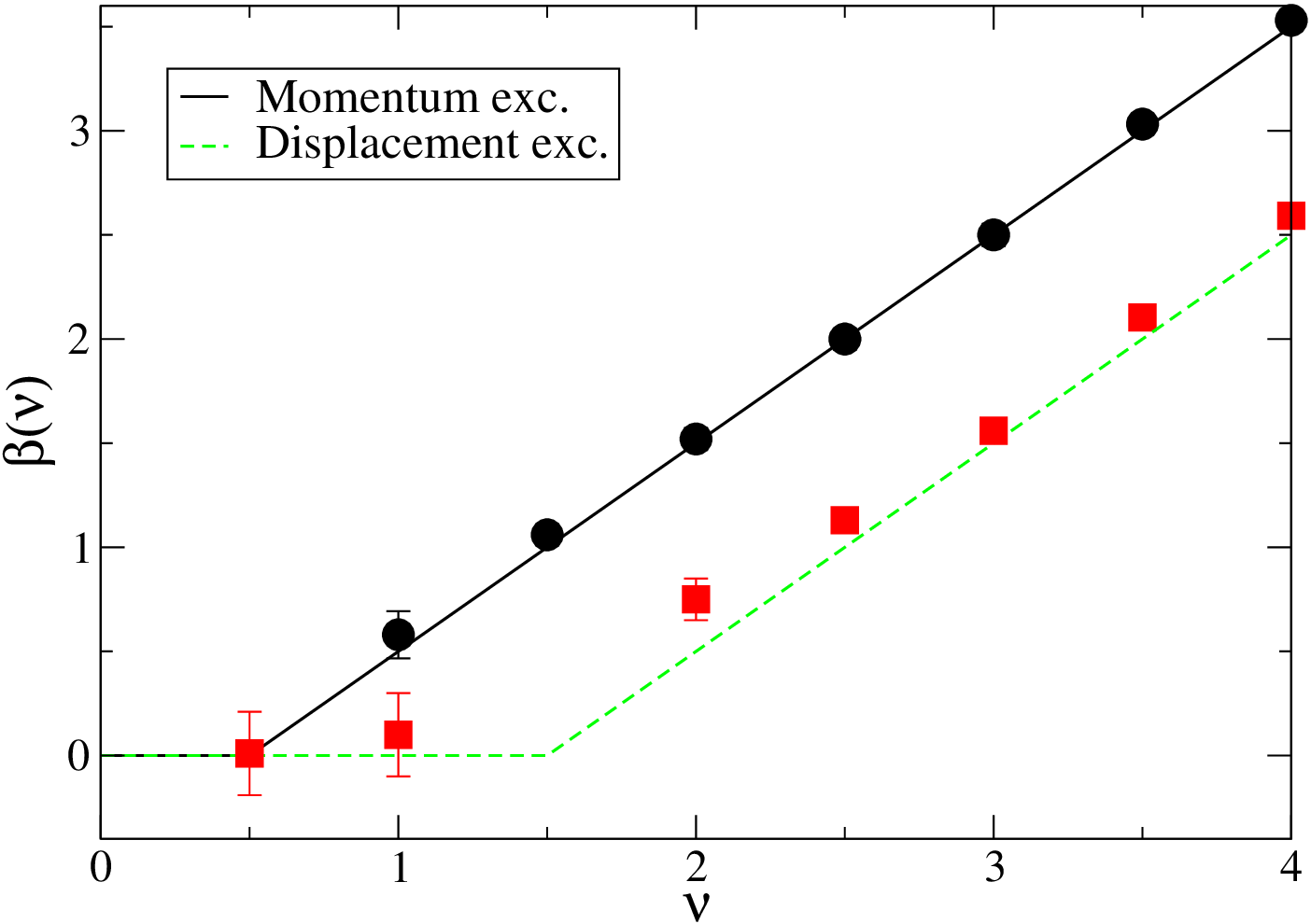}
\end{center}
\caption{(Color online) Comparison between
the exponents measured numerically for momentum 
and displacement excitations (full circles and squares 
respectively) and the analytical result Eq.~(\ref{mnu}), same parameters
as in Fig.~\ref{fig6}. The exponents are evaluated by
a power-law fit. Error bars are estimated from
the fluctuations of the (discrete) logarithmic derivative
${\Delta \log \langle m_\nu(t)\rangle}/{\Delta\log t}$
and are reported only when larger than symbols' size.
}
\label{fig6bis}
\end{figure}

For $\nu=\eta-1$ there is a logarithmic divergence of $\langle
m_\nu (t) \rangle$ with time. As demonstrated in  Fig.
\ref{fig6bis}, the numerical values of $\beta(\nu)$ are in
excellent agreement with Eq.~(\ref{mnu}). This also implies that
the contribution of the traveling peaks is not relevant as
implicitly assumed in the derivation of Eq.~(\ref{mnu}).

This result is consistent with the values that could be inferred
by Datta and Kundu \cite{dk}. Indeed, they predict $\beta(2)=1/2$ and
$\beta(2)=3/2$, respectively, for a displacement and momentum
excitation. Notice that, if one looks only at $m_2(t)$ one would
incorrectly conclude that the two cases would correspond to
sub- and superdiffusive behavior respectively. A full analysis of
the spectrum of moments and of the wavefront shape is necessary to
assess the real nature of dynamics.

\section{ENERGY PROFILE: ANHARMONIC CASE}

In this section we will investigate numerically the $n$-dependence
of the energy profile averaged over the disorder in the presence
of anharmonicity. Particularly, we will check whether its
tails can be described by those of the harmonic chain. As a model
we have chosen the Fermi-Pasta-Ulam (FPU) chain with cubic
nonlinear force
\begin{eqnarray}
m \ddot u_n &=& K_n(u_{n+1}-u_n) - K_{n-1}(u_n-u_{n-1}) \nonumber\\
&&+G(u_{n+1}-u_n)^3-G(u_{n-1}-u_n)^3. \label{fpu}
\end{eqnarray}
It reduces to the harmonic chain for $G=0$. For simplicity, we
considered the case of uniform nonlinear coupling $G$
($G=1$ in the following).

The analysis of the previous section shows that the behavior of
the harmonic chain follows all the expected features. Which
influence of the anharmonicity do we expect? If the initially
localized energy would spread completely it would be $e_n(t),
\longrightarrow 0$ for $ t \rightarrow \infty$, for all $n$. For
incomplete spreading, however, $\langle e_n(t) \rangle$ for $t$
large enough should decay by the power laws Eqs.~(\ref{asym1}) or
Eqs.~(\ref{asym2}), again, and the amplitudes of oscillations at
sites far away from site $n_0$ of the initial excitation should
become so small that the harmonic approximation applied to those
tails should become valid.

A detailed analysis of the effects of nonlinearity goes beyond the
scope of the present work. We thus limited ourselves to the case
of FPU with initial displacement excitation with $A=2$. We checked
that the energy is about a factor of 2 larger with respect to the
$G=0$ case meaning that the nonlinear part of the potential is
sizeable. We considered the usual definition of $e^{(\rm pot)}_n$
where $K_i[u_{i+1}-u_i],$ $i=n-1,n$ in Eq.~(\ref{pot}) is replaced
by $V'_i (u_{i+1}-u_i)$ with $V_i(x)=K_i x^2/2+Gx^4/4$. As for the
harmonic case, we performed the average over disorder at three
different times. 

\begin{figure}[ht]
\begin{center}\leavevmode
\includegraphics[width=0.45\textwidth,clip]{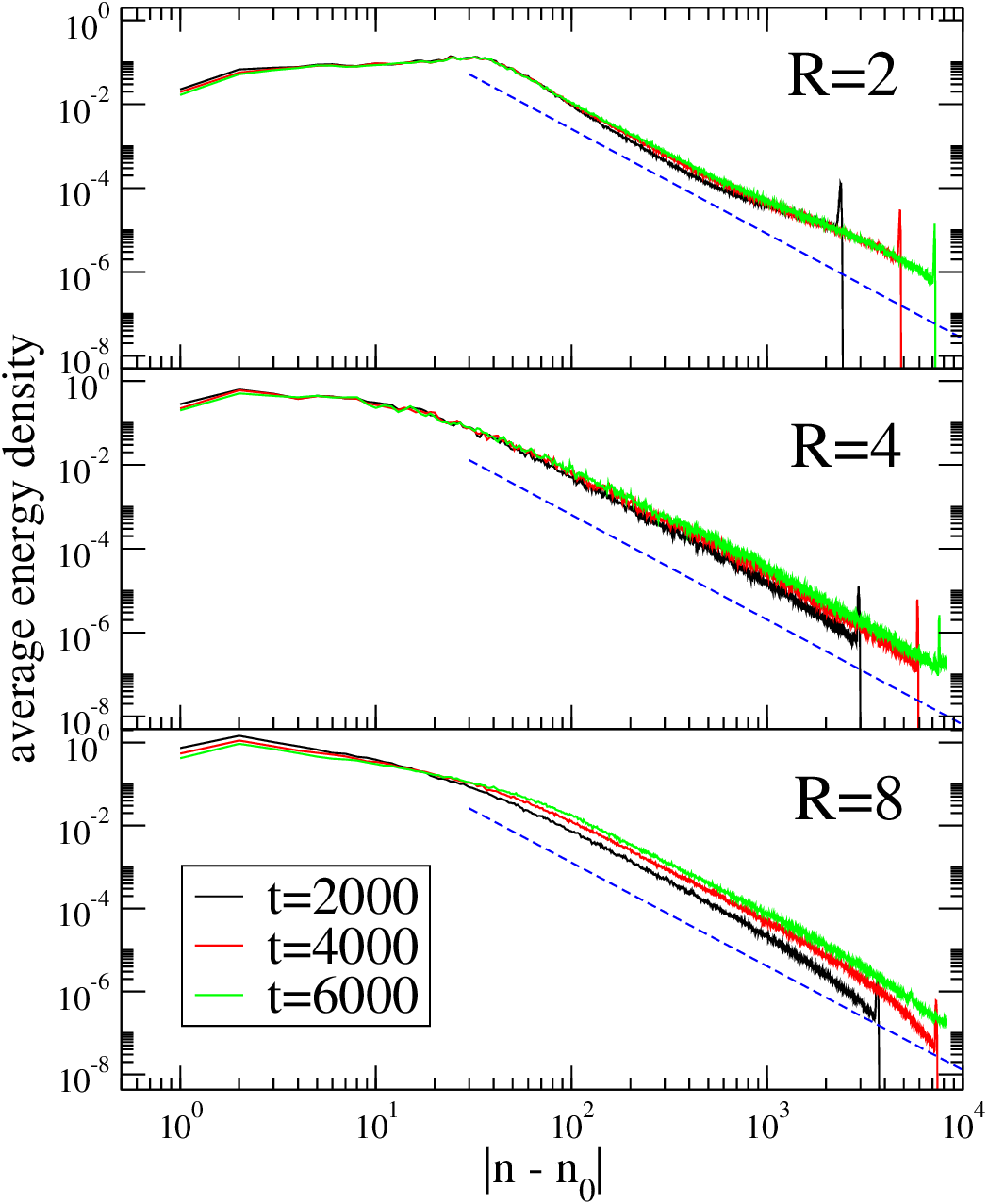}
\end{center}
\caption{(Color online) FPU model: Disordered-average
energy profile at three different times
averaged over $10^3$ realizations of the disorder and for different
disorder strengths $R$. Chain of
$N=8192$ particles with displacement excitation $A=2$.
For comparison, the predicted power law decay for the harmonic
chain, Eq.~(\ref{asym1}), is also drawn (dashed lines).
}
\label{fig7}
\end{figure}

The average energy profiles for three different
disorder strengths are reported in Fig.~\ref{fig7}. The profiles
still show a pretty slow decay, reminiscent of the harmonic case.
From Fig. \ref{fig7} we first observe that the convergence
of $\langle e_n (t) \rangle $ to a limiting profile at $t=\infty$
becomes slower for larger disorder strength $R$, i.e. for shorter
localization lengths. Second, whereas the profile for $R=4$ and
the largest time $t=6000$ can be satisfactorily fitted by the
power law Eq.~(\ref{asym1}), this is less obvious for $R=2$ and
$R=8$. For $R=2$ and $|n-n_0| > 1000$ the profile is practically
time independent for $t \geq 2000$. But in contrast to the
harmonic case (see Fig.~\ref{fig2}) it does not reveal the
asymptotic power law $|n-n_0|^{-5/2}$, although the data suggest
that this may happen for $|n-n_0| \geq 3000$. For $R=8$ the
profile follows that power law for $100 < |n-n_0| < 1000$, i.e.
for about a decade, but deviates for $|n-n_0|>1000$. However,
comparing this profile for the three different values for $t$
hints that the range of the power law decay may increase with
increasing $t$. In addition, the profiles display some
form of weak "broadening" of the tails indicating that some energy
is indeed slowly propagating. 

\begin{figure}[ht]
\begin{center}\leavevmode
\includegraphics[width=0.45\textwidth,clip]{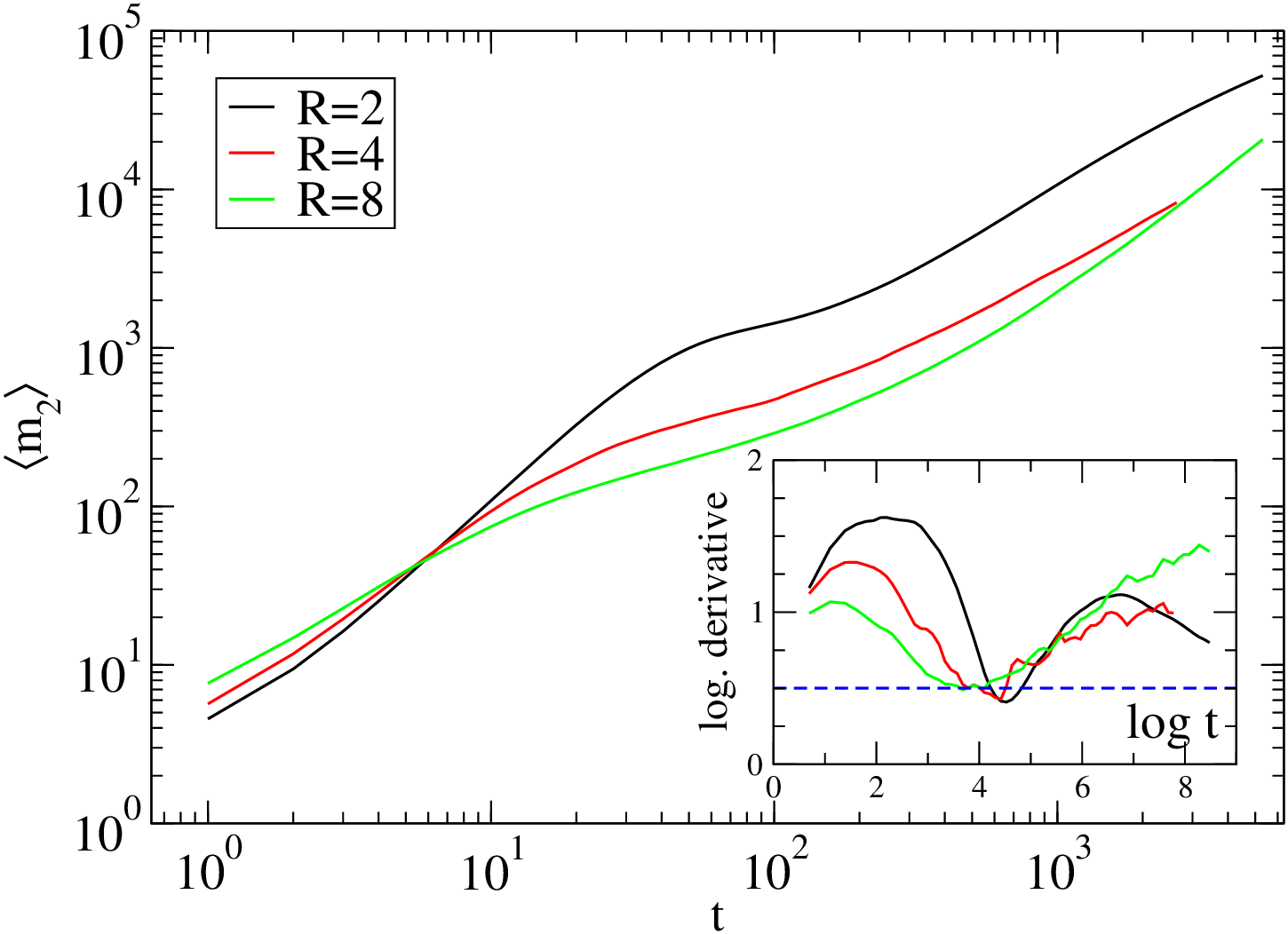}
\end{center}
\caption{(Color online) FPU model: Evolution of
$\langle m_2(t) \rangle$  for
$R=2, 4, 8$ (top to bottom respectively)
same parameters as in Fig.~\ref{fig7}.
The inset reports the (discrete) logarithmic derivative
${\Delta \log \langle m_2(t)\rangle}/{\Delta\log t}$
of the data versus $\log t$. For comparison, the value 
for the harmonic chain $\beta(2)=1/2$ is also drawn
(dashed horizontal line). 
}
\label{fig8}
\end{figure}

As a consequence, the disorder
averaged moments $\langle m_\nu (t) \rangle$ do not
display a convincing scaling with time. Even for statistically
accurate data as the one in Fig.~\ref{fig7}, the effective
exponents (as measured for example by the logarithmic derivatives
of  $\langle m_\nu (t) \rangle$) display sizeable oscillations
which are well outside the range of the statistical 
fluctuations (see Fig.~\ref{fig8}). Similar results are obtained
for momenta of different order (not reported).

We  may thus argue that, at least in the considered parameter
range, the nonlinear case has a core which remain almost localized
(in a similar way as the harmonic case) but in addition there must
be a small propagating component. The fraction of such propagating
component increases upon increasing the energy and/or
nonlinearity. As a consequence, with the data at hand it is
impossible to draw definite conclusions on the nature of the
spreading process.

\section{SUMMARY AND CONCLUSIONS}

The relaxation of an initially localized excitation in a
translationally invariant chain of particles has been studied for
harmonic and anharmonic nearest neighbor couplings. The main focus
has been on the energy profile $\langle e_n(t) \rangle$, the
moments $\langle m_\nu (t)\rangle$, both averaged over the
disorder, and the relation between the asymptotic $t$-dependence
of $\langle m_\nu (t) \rangle$ with the asymptotic $n$-dependence
of $\langle \overline{e_n (t)} \rangle$. As far as we know, this
has neither been explored for the anharmonic nor for the harmonic
case due to the lack of analytical knowledge of $\langle
\overline{e_n (t)}\rangle$ for $|n-n_0| \rightarrow \infty$.

For the harmonic model we succeeded to determine analytically the
disorder and time averaged energy profile $\langle
\overline{e_n(t)}\rangle$ for a displacement and a momentum
excitation at site $n_0$ and initial time $t=0$. Whereas $e_n(t)$
is a quasiperiodic function which does not converge for $t
\rightarrow \infty$ we have argued that $\langle e_n(t)\rangle$
converges for $t \rightarrow \infty$. In that case $\langle
\overline{e_n(t)}\rangle $ gives the limiting profile averaged
over the disorder. The analytical calculation yields a power law
decay
\begin{equation} \label{powlaw}
\langle \overline{e_n(t)}\rangle \cong C(R) |n-n_0|^{-\eta}
\end{equation}
for $1 \ll |n-n_0| \ll N$, in case the system is finite. The
exponent $\eta$ and the prefactor $C(R)$ depend on the type of
excitation. For a displacement and momentum excitation we have
found $\eta =5/2$ and $\eta=3/2$, respectively, in good agreement
with the numerical values. This agreement also holds for the
analytical and numerical results for the $R$-dependence of $C(R)$,
except for the two smallest values of $R$ in case of a momentum
excitation. Accordingly our assumption $\xi_2(\omega) \sim
\xi(\omega)$ for $\omega \rightarrow 0$ is supported. From this
proportionality it also follows that $C(R)$ diverges at $R=1$, the
no-disorder limit. The power law decay, Eq.~(\ref{powlaw}), 
originates from the gapless excitation spectrum of the Anderson
modes. It is the consequence of the translational invariance of
model ~(\ref{ham}). Destruction of this invariance by adding, e.g.
an on-site potential like in the KG model generates an energy gap.
The corresponding localization length at the lowest eigenfrequency
will not diverge anymore, and therefore the energy profile will
decay exponentially for $|n-n_0| \rightarrow \infty$.
However, we stress that {\it any} lattice model without an
external potential has to be invariant under arbitrary
translations. This implies a gapless spectrum which is the origin
of the power law decay of the profile.

The power law decay of $\langle \overline{e_n(t)}$ has
remarkable consequences on the asymptotic $t$-dependence of the moments. 
If we use Eq.~(\ref{powlaw}) to calculate the time and disorder
averaged $\nu$-th moment
we get for displacement and momentum excitation:
\begin{equation} \label{avmom1}
\langle \overline{m_\nu(t)} \rangle = \infty
\end{equation}
for {\it all} $\nu \geq 2$; note that, for instance, 
$\langle \overline{m_1(t)} \rangle$ is
finite for a displacement, but not for a momentum excitation. The
result, Eq.~(\ref{avmom1}), implies that the disorder averaged
moment $\langle m_\nu (t) \rangle$ must diverge with time,
although the initial local energy excitation does not spread
completely. This power law divergence of $\langle m_\nu (t)
\rangle$ with time is clearly supported by the numerical result
for $\nu \geq 2$ and $\nu \geq 1$ for the displacement and
momentum excitation, respectively. As a matter of fact,
consideration of $m_2(t)$ alone is not sufficient to conclude that
the energy diffuses. This is one of the main messages of the paper.

The analytically exact result for $\langle \overline{e_n(t)}
\rangle$ in case of harmonic interactions also allows to check how
far the tails of an anharmonic chain, where the average
displacements become arbitrary small, can be described by the
tails of the harmonic system Although no definite conclusion can
be drawn, we have found evidence for a crossover of the energy
profile of the anharmonic to that of the harmonic chain. However,
for the weakest and strongest strength of disorder this crossover
seems to occur for $|n-n_0|>3000$ and for $t > 6000$,
respectively. This may be explained as follows. The localization
length $\xi_2(\omega)$ is large for weak disorder. Since
$|n-n_0|/\xi_2 (\omega)$ enters into the calculation of the
disorder averaged profile (see Eq.~(\ref{avprof2})) the asymptotic
power laws, Eqs.~(\ref{asym1}) and ~(\ref{asym2}), occur at larger
values of $|n-n_0|$. For large disorder, $\xi_2 (\omega)$ is
small. But the time scale for tunneling processes responsible for
the energy propagation increases significantly due to an increase
of the potential barriers. Therefore the convergence to a limiting
profile is much slower which is exactly what we observed (see
Fig.~\ref{rel}). The increase of the localization length for weak
disorder and the increase of the relevant time scale of the
relaxation for strong disorder probably are also the reasons for
the absence of a convincing scaling of the moments $\langle m_\nu
(t) \rangle$. In order to test this, one has to increase both, the
number of particles and the simulation time significantly.
Requiring a similar good statistic of the data this has not been
possible so far within the available CPU time. If it is true that
the asymptotic energy profile agrees with that of the harmonic
chain this would imply that the moments $\langle m_\nu (t)
\rangle$ for the anharmonic model for $\nu \geq 2$ diverge with
time, as well, although the energy does not spread completely.

From our results, it is nonetheless clear that the interplay of
localized and almost-extended modes leads to a nontrivial decay of
wavepackets amplitudes and this must be taken into account when
dealing with the nonlinear case.

\acknowledgments

We thank S. Flach for stimulating discussions and gratefully
acknowledge the MPI-PKS Dresden for its hospitality and financial
support. SL is partially supported by the CNR
\textit{Ricerca spontanea a tema libero} N. 827
\textit{Dinamiche cooperative in strutture quasi
uni-dimensionali}.

\end{document}